\documentclass[useAMS,usenatbib]{mn2e}
\usepackage{graphicx}

\def\be{\begin{equation}} 
\def\ee{\end{equation}}

\def\HI{\hbox{H~$\scriptstyle\rm I\ $}}

\def\gsim{\lower.5ex\hbox{\gtsima}} 
\def\lsim{\lower.5ex\hbox{\ltsima}} \def\gtsima{$\; \buildrel > \over 
\sim \;$} \def\ltsima{$\; \buildrel < \over \sim \;$} \def\prosima{$\; 
\buildrel \propto \over \sim \;$} \def\gsim{\lower.5ex\hbox{\gtsima}} 
\def\lsim{\lower.5ex\hbox{\ltsima}} 
\def\simgt{\lower.5ex\hbox{\gtsima}} 
\def\simlt{\lower.5ex\hbox{\ltsima}} 
\def\simpr{\lower.5ex\hbox{\prosima}} \def\la{\lsim} \def\ga{\gsim} 
  
 \def\gtsima{$\; \buildrel > \over \sim \;$} 
\def\ltsima{$\; \buildrel < \over \sim \;$} 
\def\gsim{\lower.5ex\hbox{\gtsima}} 
\def\lsim{\lower.5ex\hbox{\ltsima}} 
\def\simgt{\lower.5ex\hbox{\gtsima}} 
\def\simlt{\lower.5ex\hbox{\ltsima}} 
\def\simpr{\lower.5ex\hbox{\prosima}} 
\def\la{\lsim} 
\def\ga{\gsim}

\def\E3{{\cal E}_{\rm g}^{III}}

\def\Msun{\rm M_\odot}

\title[Fundamental metallicity relation]{The physics of the fundamental metallicity relation } 
\author[Dayal et al.]{Pratika Dayal$^{1}$\thanks{E-mail:prd@roe.ac.uk (PD)}, Andrea Ferrara$^{2}$, James S Dunlop$^{1}$  \\ 
$^{{1}}$ SUPA\thanks{Scottish Universities Physics Alliance}, Institute for Astronomy, University of Edinburgh, Royal Observatory, Edinburgh, EH9 3HJ, UK \\
$^{2}$ Scuola Normale Superiore, Piazza dei Cavalieri 7, 56126 Pisa, Italy}

\begin{document} 
 
\date{} 
 
 
\maketitle 
 
\label{firstpage} 
\begin{abstract} 
We present a simple, redshift-independent analytic model that explains the local Fundamental Metallicity Relation (FMR), 
taking into account the physical processes of star formation, inflow of metal-poor intergalactic medium (IGM) gas, and the outflow of metal rich 
interstellar medium (ISM) gas. We show that the physics of the FMR can be summarised as follows: for massive galaxies with stellar 
mass $M_* \geq 10^{11} \Msun$, ISM metal enrichment due to star formation is compensated by inflow of metal poor IGM gas, 
leading to a constant value of the gas metallicity with star formation rate (SFR); outflows are rendered negligible as a result of the large 
potential wells of these galaxies. On the other hand, as a result of their smaller SFR, less massive galaxies produce less heavy elements that
are also more efficiently ejected due to their shallow potential wells; as a result, for a given $M_*$, the gas metallicity decreases with SFR. For 
such galaxies, the outflow efficiency determines both the slope, and the knee of the metallicity-SFR relation. Without changing any parameters, this simple model is also successfully matched to the gas fraction - gas metallicity relation observed for a sample of about 260 nearby galaxies.

\end{abstract}

\begin{keywords}
 galaxies:high-redshift - fundamental parameters - evolution - abundances - stellar content 
\end{keywords}

\section{Introduction}

Observationally, the stellar mass and the gas phase metallicity are two of the most fundamental properties of galaxies, reflecting the amount of 
baryons locked up in stars, and the amount of metals present in the interstellar medium (ISM) gas, respectively. They also provide insights into a number of parameters pertinent for understanding galaxy evolution including the stellar initial mass function (IMF) which determines the amount of metals produced and returned by stars to the ISM, the star formation history which determines the total stellar mass and metal content, and the amount of metal rich gas ejected/mass of metal poor gas accreted. The existence of a mass-metallicity relation has now been observationally established by many authors including \citet{garnett2002}, \citet{tremonti2004}, \citet{cowie-barger2008} and \citet{perez-montero2009}. The origin of this relation, however,
has been widely debated and a number of different explanations have been proposed for its existence, including: the ejection of metal rich 
gas \citep[e.g.][]{garnett2002,spitoni2010}, a dependence of the star formation rate (SFR) on the galaxy mass \citep[e.g.][]{calura2009}, infall of metal 
poor intergalactic medium (IGM) gas into the galaxy \citep[e.g.][]{dekel2009a,mannucci2009}, and a balance between ISM metal enrichment due to star formation and dilution due to infall of metal poor IGM gas \citep{finlator-dave2008}.

Recently, using observations of galaxies at $0 < z < 4$, \citet{mannucci2010} have shown that the mass-metallicity relation arises as a 
consequence of a more fundamental 3D relation, referred to as the fundamental metallicity relation (FMR), relating the stellar mass, gas-phase metallicity and the SFR \citep[see also][]{lara-lopez2010}; indeed, if infall is the main driver for star formation, and it is star formation that drives outflows, such a relation is only to be expected. 

\begin{figure*}
\begin{center}
\center{\includegraphics[scale=0.9]{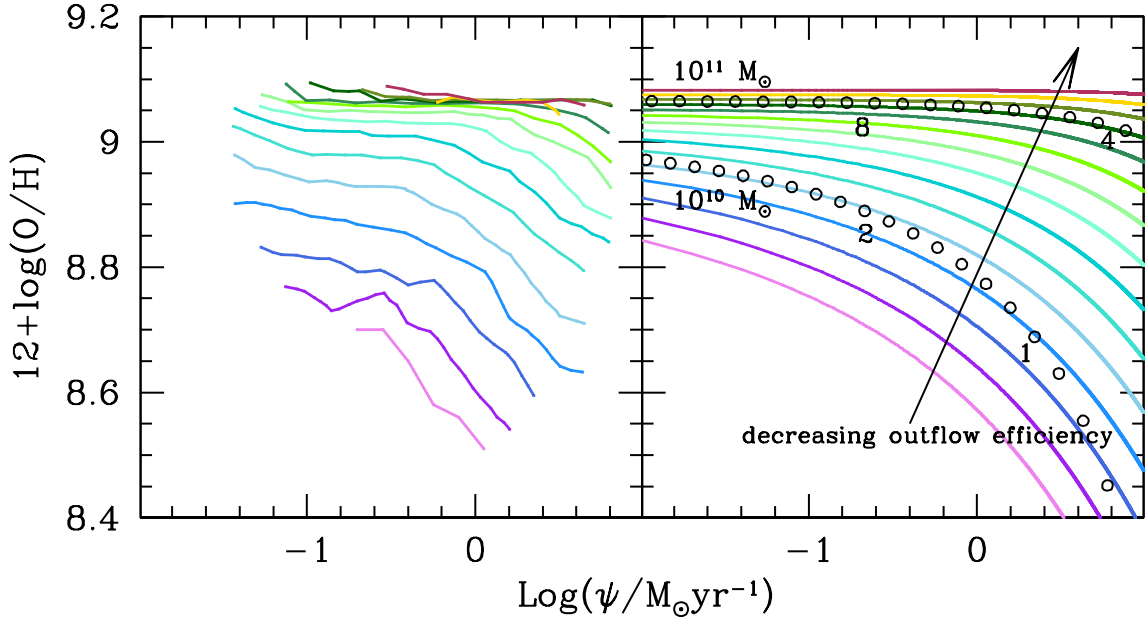}} 
\caption{Oxygen abundance versus SFR for galaxies of different stellar masses, in the range $M_*=10^{9.25+0.15j} \Msun$, with $j=0,..,14$, from the bottom to the top curves, in both panels. The left panel shows the oxygen abundance-SFR relation inferred observationally by \citet{mannucci2010} for local SDSS galaxies while the right panel shows the theoretical results. In the right panel, points show the evolutionary tracks of two representative galaxies with initial gas masses of $10^{10,11} \Msun$ respectively, and numbers along the track indicate the values of $\tau =\epsilon_* t$ at selected ages. }
\label{analytic}
\end{center}
\end{figure*}

In this work, our aim is to build a simple and yet physically consistent analytic model that explains the observed FMR, taking into account the relevant physical processes of gas inflow/outflow, star formation and metal production. The main advantage of the model presented here lies in the fact that its results are essentially independent of the redshift considered for $z \leq 6$, for reasons that are explained in what follows. As an excellent validation of the model, we show that it also reproduces the gas fraction and metallicity relation observed for about 260 galaxies in the local Universe.

\section{Basic physical interpretation}
\label{basics}
As explained above, our first aim is to build the {\it simplest} model that can account for the observed FMR trends. The observable quantities determining the FMR of a given galaxy, i.e. stellar mass ($M_*$), SFR ($\psi$), and oxygen abundance (often used as a proxy for metallicity), 
$X = M_O/M_g$, where $M_O$ and $M_g$ are the galaxy oxygen and total gas mass, respectively, are related by a simple set of evolutionary ordinary 
differential equations:
\begin{equation}
\label{evoeq1}
\frac{dM_*}{dt}\equiv \psi = \epsilon_* M_g
\end{equation}
\begin{equation}
\label{evoeq2}
\frac{dM_g}{dt} = -(1-R)\psi + (a-w) \psi
\end{equation}
\begin{equation}
\label{evoeq3}
M_g\frac{dX}{dt} = y(1-R) \psi - aX\psi,
\end{equation}
where $\epsilon_*^{-1}$ is the star formation timescale and the SFR is assumed to be proportional to the gas mass. The two constants, $R$ and $y$, represent the returned fraction from stars and the the yield per stellar generation, respectively, and are dependent on the IMF; whenever numerical values are required, we use $(R, y) = (0.79, 0.0871)$ consistent with a Salpeter IMF for a lower (upper) mass limit of 1 (100) $\Msun$. We assume that both the outflow, $w(M)\psi$, and infall, $a(M)\psi$, rates are proportional to the SFR, where $M$ is the total (dark+baryonic) galaxy mass; while the former assumption is easy to understand, the latter has been made to obtain the simplest solution to the equations described above. We further pose that the metallicity of the infalling gas is $X_i \ll X$.
Simple algebraic manipulation of the above equations gives the general solution 
\begin{equation}
\label{solution}
X(\psi) = \frac{y(1-R)}{a}\left(1-\mu^{-\alpha}\right),
\label{master}
\end{equation}
where $\mu=M_g/M_{g0}  = \psi [\epsilon_*M_{g0}]^{-1}$. The initial gas mass can be expressed as $M_{g0}\simeq M_*(1+w-a)+M_g = (\Omega_b/\Omega_M)M$, and $\alpha=a[R-1+a-w]^{-1}$. We now interpret the physical meaning of the above relation. 

\begin{figure*}
\begin{center}
\center{\includegraphics[scale=0.9]{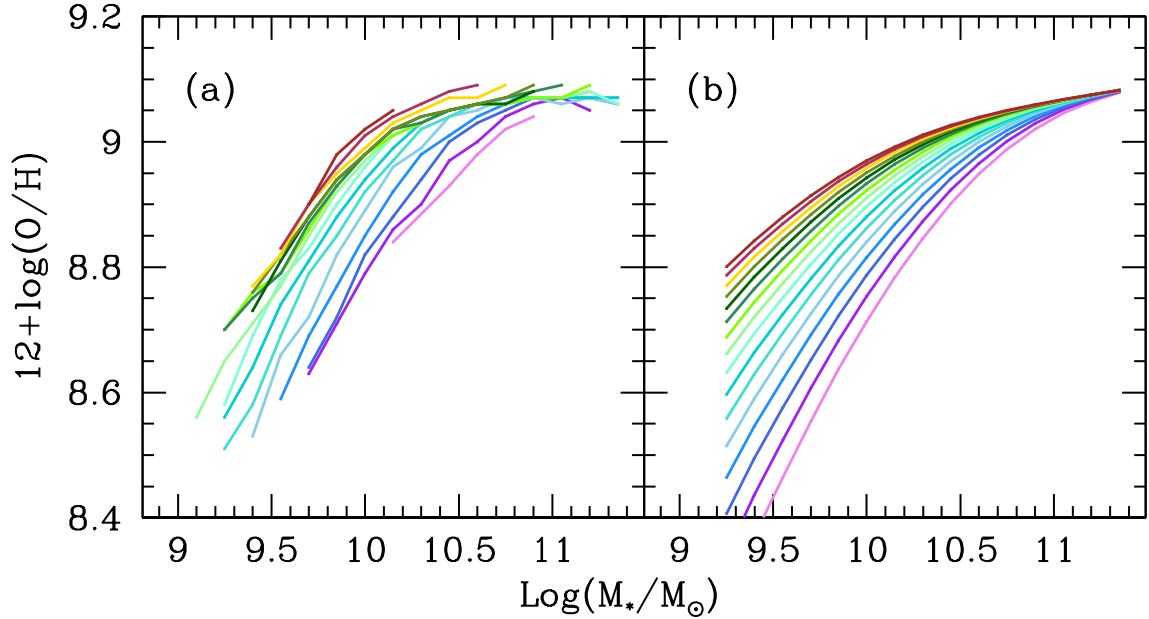}} 
\caption{Oxygen abundance versus $M_*$ for galaxies of different SFR, in the range $\psi=10^{0.8-0.15j} \Msun$, with $j=0,..,15$, from the bottom to the top curves, in both panels. The left panel shows the oxygen abundance-$M_*$ relation inferred observationally by \citet{mannucci2010} for local SDSS galaxies while the right panel shows the theoretical results. }
\label{metg_fnms}
\end{center}
\end{figure*}

Firstly, we note that reproducing the observed decreasing trend of $X(\psi)$ for the smallest galaxies requires $\alpha<0$, i.e. the condition $a < w +(1-R)$ has to be satisfied. Note that this condition also ensures that $\mu^{-\alpha} \le 1$ at any time. In addition, if $\mu \ll 1$, as is the case for large galaxies in which most of the gas has been turned into stars,  $X(\psi) \approx y(1-R)a^{-1} =$ constant, as indeed observed. This property can then be used to normalize the value of $a$ for the largest galaxies to the observed oxygen abundance, $X^{obs}$ by requiring $a = y(1-R)/X^{obs}$. Physically, the larger the value of $a$, smaller is the value of $X$, i.e. the metallicity of the galaxy is diluted by low metallicity infalling gas. We note that the no-infall case ($a=0$) would not admit solutions in which metallicity is independent of the SFR; such constancy implies that some infall is required.  

Outflows affect the solution through the term $\mu^{-\alpha}$ in two ways: (a) by setting the position, $\psi^*$, of the knee in the $X(\psi)$ curve, above which the metallicity starts to drop at a fixed $M_*$; this is determined by the condition $\mu = 0.5$, or $\psi^* = \epsilon_* (1+w-a) M_*$. Thus, for large outflow efficiencies ($w$), the knee shifts to lower values of $\psi$; at fixed $w$, less massive galaxies deviate from a constant $X(\psi)$ relation at lower $\psi$ values. (b) The outflow efficiency $w$ also regulates the slope of the curve through the power $\alpha$: as $w$ is increased the slope becomes steeper. The combined effect is that, for a given value of $\psi$, low-mass galaxies are more metal poor than massive ones. 

By fitting the $X(\psi)$ curves for various $M_*$ to the \citet{mannucci2010} data (panel a of Fig. \ref{analytic}), we can also tune the dependence of the only two free parameters of the model: $a(M)$ and $w(M)$. In a previous paper \citep[see Fig. 6,][]{dayal2009}, using state-of-the-art cosmological SPH simulations, we have shown that $M_* \propto M$; as a result, we can assume that the two quantities track each other. We find that the best fit to the data require 
\begin{equation}
\label{afit}
\ln a = -0.43 - 0.05 \ln (M_*/10^{10.75}) \Msun,
\end{equation}
\begin{equation}
\label{wfit}
\ln w = 1.76 - 0.33 \ln (M_*/10^{9.0}) \Msun,
\end{equation}
Hence, the accretion efficiency is essentially independent of the galaxy mass, implying that the accretion rate scales almost linearly with $M$. The ejection efficiency is small in large galaxies ($w\approx 1.0$ for $M_*=10^{11.35} \Msun$) and rapidly increases with decreasing $M_*$, reaching $w=4.8$ for the smallest ones ($M_*=10^{9.25} \Msun$). It is interesting to note that the mass dependence we find, $w \propto M_*^{-1/3} \propto v_e^{-1}$, is the one expected if the outflow is {\it momentum-driven}.

The $X-\psi$ relation (solid curves) as a function of the $M_*$ is shown in panel (b) of Fig. \ref {analytic}, along with the evolutionary tracks (points) of two representative galaxies whose initial gas mass is $M_{g0}=10^{10, 11}\Msun$ (time runs from right to left along the open circles). Galaxies start from low $X$, low $M_*$, and high $\psi$ values and then move towards higher metallicities as their stellar mass increases. However, depending on their mass they move at different velocities along the track: the most massive observed galaxies are very evolved with a specific age (i.e. in units of the star formation timescale $\epsilon_*^{-1}=0.62$ Gyr)  $\tau =\epsilon_* t \simgt 4$, and move essentially along constant metallicity curves; smaller objects are younger ($1 < \tau < 2$) and are still gently building up their metal content. 

\section{Projections of the FMR}
\label{proj}
We now compare our model against a second projection of the three-dimensional FMR: that relating $X$ and $M_*$ for different $\psi$ bins. Observationally, \citet{mannucci2010} find that while $X$ decreases with increasing $\psi$ values for $M_* \leq 10^{11} \Msun$, it is independent of $\psi$ for larger $M_*$ values, as shown in panel (a) of Fig. \ref{metg_fnms}. These same trends are seen from the theoretical FMR presented in panel (b) of Fig. \ref{metg_fnms}, where the theoretical and observed results are in excellent agreement, both in terms of the amplitude and the slope of the relation. This is almost to be expected given that, as explained in Sec. \ref{basics}, for a given $M_* <10^{11} \Msun$, an increase in $\psi$ leads to an increase in the ejection efficiency, $w$, thereby reducing the dust content of these galaxies. On the other hand, for systems that have already assembled a stellar mass $M_* >10^{11} \Msun$, outflows are rendered unimportant as a result of the large potential well characterising these objects and $X$ becomes independent of $M_*$ since the ISM metal enrichment from star formation is compensated by the inflow of metal poor IGM gas. 

Recently, using a magnitude and volume limited sample of 260 nearby late-type galaxies, \citet{hughes2012} have presented results relating their gas metallicity, stellar mass and gas content; the gas content refers to the neutral hydrogen (\HI) content, inferred using the observed 21cm emission. These authors show that for a given value of $M_*$ ranging between $10^{9-11.4}\Msun$, galaxies with lower \HI gas fractions, $f_{HI}=M_{HI}(M_{HI}+M_*)^{-1}$,  show higher values of $X$. This result is found to hold true both for galaxies that have a normal (i.e. expected) \HI content, and galaxies that are `\HI-deficient', i.e. contain $\geq 70$\% less \HI mass than is expected for isolated objects of the same morphological type and optical size.  

\begin{figure}
\begin{center}
\center{\includegraphics[scale=0.45]{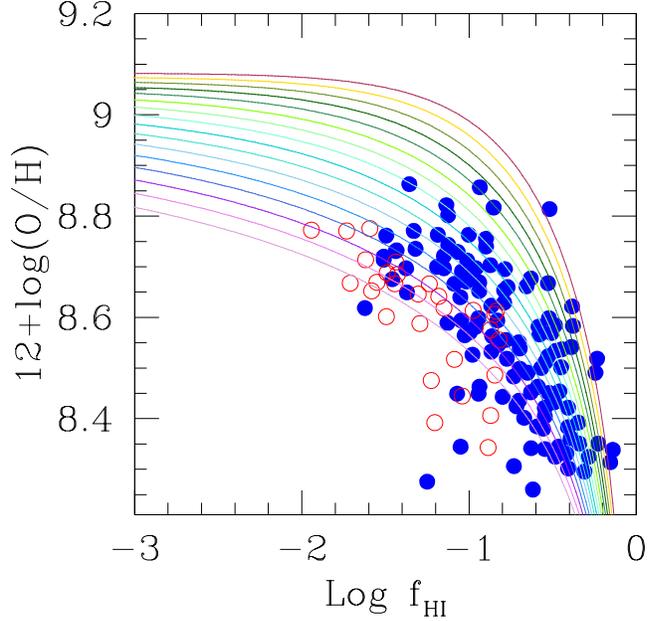}} 
\caption{Oxygen abundance versus the \HI gas fraction $f_{HI} = M_{HI}(M_{HI}+M_*)^{-1}$: the empty (solid) points show the values inferred for \HI-deficient (normal) galaxies by \citet{hughes2012} using a volume limited sample of 260 nearby late-type galaxies. Solid lines show the theoretical results for different stellar masses, in the range $M_*=10^{9.1+0.15j} \Msun$, with $j=0,..,15$, from the bottom to the top curves.}  
\label{metg_fnh1}
\end{center}
\end{figure}

To compare to these results, we rewrite Eq. \ref{master} in terms of the \HI mass, $M_{HI}$, instead of the total gas mass, $M_g$, as 
\begin{equation}
X(f_{HI}) = \frac{y(1-R)}{a}\bigg\{1-\bigg(\frac{\mu_2}{\mu_2+M_*(1+w-a)}\bigg)^{-\alpha}\bigg\},
\label{fh1}
\end{equation}
where $\mu_2 = f_{HI}M_*[0.3(1-f_{HI})]^{-1}$ and all the other symbols have meanings and values as in Sec. \ref{basics}. The factor 0.3 in $\mu_2$ arises because we assume about 30\% of the total gas content of any galaxy to be in the form of \HI, i.e. $M_{HI}/M_g = 0.3$, using the fiducial values inferred observationally for the Milky Way \citep{ferriere2001}. We then calculate the $X-f_{HI}$ relation for the stellar mass range observed by \citet{hughes2012}, using quantised values of $M_*$ spaced equally by 0.15 dex between $10^{9.1}$ and $10^{11.35}\Msun$. We find both the amplitude and the slope of the theoretical results are in excellent agreement with the observations, as shown in Fig. \ref{metg_fnh1}. This is a heartening success of the model, given that once the two free parameters of our model ($a,w$) are fixed by comparing to the observations of \citet{mannucci2010}, we have no free parameters left to match to the data set collected by \citet{hughes2012}, and that both these data sets have been collected using very different observational techniques. 

For an $M_*$ range spanning about two orders of magnitude, for a given $M_*$, $f_{HI}$ decreases with increasing $X$. Alternatively this implies that for a given stellar mass, galaxies that have assembled a larger stellar mass are also more metal enriched. However, the slope of the $f_{HI}-X$ relation steepens with increasing $M_*$ values as shown in Fig. \ref{metg_fnh1}: galaxies with low $M_*$ values are slowly building up both their stellar mass and metal content and the gas fraction decreases as gas is both consumed in star formation and lost in outflows. On the other hand, more massive galaxies that start with larger gas reservoirs consume gas in an initial burst after which $\psi$ decreases and $X$ flattens as a result on the balance between metal production from star formation and the dilution due to infalling IGM gas. 

Finally, although it is a success of the model that it is able to explain the $X-f_{HI}$ relation for the bulk of the galaxies observed by \citet{hughes2012} without needing to invoke any free parameters, there are a few outliers with $X$ and $f_{HI}$ values lower than those predicted by our model. The reasons for this discrepancy are now discussed: we have used a fiducial value of $M_{HI} = 0.3 M_g$, inferred for the Milky Way. In reality however, this ratio depends on a number of complex, interlinked and poorly understood processes including star formation, the temperature, density and ionization distribution in the ISM and the effects of turbulence. Further, the `HI-deficient' galaxies observed by \citet{hughes2012} have, by definition, a lower \HI content than expected for a galaxy of their size and type.

\section{Conclusions and discussion}
\label{conc}
We have presented a very simple model that provides a straightforward explanation of the 3D FMR relating the stellar mass ($M_*$), SFR ($\psi$) and gas metallicity ($X$) observed by \citet{mannucci2010} in the local Universe. Starting from a galaxy that has already assembled its gas mass after a transient gas-accretion phase, assuming the metallicity of the infalling IGM gas to be much lower than that of the ISM gas, and assuming both the inflow and outflow rates to be proportional to the $\psi$, we calibrate the only two model free parameters (the inflow and outflow rates) to the local FMR data. Using this model, we show that both inflows and outflows are required to explain the observations, although their importance is mass dependent: since we assume the inflow rate to be proportional to $\psi$ (and hence $M_*$), galaxies with the largest $M_*$ accrete the largest amount of metal-poor IGM gas, which is balanced by the larger amount of metals produced due to star formation; this, coupled with the negligible outflow rates due to their large potential wells, makes the $X$-$\psi$ relation essentially constant for galaxies with $M_* \geq 10^{11} \Msun$. On the other hand, due to their smaller values of $\psi$, less massive galaxies produce less metals. However, as a result of their smaller potential wells, outflows lead to a loss of metal-rich ISM gas so that at a given $M_*$, galaxies with the largest $\psi$ are the most metal poor; alternatively, for a given $\psi$, smaller galaxies are more metal poor than massive ones. Without needing to invoke any free parameters, this relation also provides an elegant explanation for the 3D relation linking $M_*,X$ and the gas content ($f_{HI}$) of nearby galaxies observed by \citet{hughes2012}: for a given $M_*$, galaxies that have converted a larger fraction of their gas mass into stars (i.e. have a lower $f_{HI}$ value) are expected to be more metal enriched. 

The strength of the simple model presented here lies in the fact that it reproduces both the locally observed FMR and $X-M_*-f_{HI}$ relations extremely well, essentially through the single, redshift-independent Eq. \ref{master}. Time (or $z$) only specifies when in the course of its evolution a galaxy reaches a given$(X, \psi)$ point in the plane. Further, in our model, the $X-M_*-\psi$ or $X-M_*-f_{HI}$ relations are shaped by the evolutionary processes taking place within the galaxy. This scenario has recently been lent strong support by the observations of \citet{hughes2012} who find that the $X-M_*$ relation is purely shaped by the internal evolution of a galaxy, with environmental effects playing only a minor role.

Finally, we comment on how our model compares to a few other existing works. In their work that presents the data, \citet{mannucci2010} have briefly discussed two possible scenarios that could lead to the FMR: (a) galaxies on the FMR are in a transient phase of high star formation and low metallicity following a period of infall of metal-poor IGM gas into the galaxy, or (b) galaxies are in a quasi-steady state where infall, star formation and gas/metal ejection occur simultaneously and outflows regulate the FMR; in this case, outflows are inversely proportional to the mass of the galaxy and increase with the SFR. As explained, our model, where the FMR arises due to an interplay between star formation, the infall of metal-poor IGM gas and the ejection of metal rich ISM gas embodies the spirit of both of the above mentioned scenarios. 

We also compare our model to that presented by \citet[][D12]{dave2012} since it is the closest in spirit to ours: although some of the conclusions reached in these two works are quite similar, there are important conceptual differences. Firstly, D12 make the {\it Ansatz} that the galaxy system is always in a stationary state, i.e. the infall rate exactly balances the gas lost in outflows and star formation; galaxies that are driven off such equilibrium due to stochastic inflows tend to be driven back towards the steady state. This is equivalent to setting ${dM_g}/{dt} = 0$ in Eq. \ref{evoeq2} above, or $a = (1-R) + w$. A posteriori, the values for the infall and outflow rate we find by fitting the $z=0$ FMR data are such that $a < (1-R) + w $, due to the large outflow rates deduced; it is important to note that in our model, both the inflow and outflow rates are related to the SFR and hence the gas mass at any given time. Further, in D12, $dX/dt=0$ is imposed on the metallicity (Eq. \ref{evoeq3}), yielding $X= y(1-R) a^{-1}$ in D12. By inspecting Eq. \ref{master}, we see that this expression approximates the relation for large galaxies with $\mu \ll 1$, but fails to catch the metallicity decrease with increasing SFR seen in smaller, outflow-dominated galaxies. Moreover, D12 impose a given external infall rate derived from numerical simulations (devised to model relatively large galaxies), while we do not make any assumption on its mass and redshift dependence; in both models though, the infall rate is proportional to $\psi$ and outflows are found to be momentum driven. Although it is not clear at this stage how the D12 model would perform when compared to the FMR data reproduced here, it would be useful to compare these two approaches in more detail. 

We conclude by discussing the main caveats: firstly, our set of time-dependent ODEs (2)-(3) require an initial condition $M_{g0} \neq 0$ in order to lead to physically meaningful solutions. This implies that our SFR always exponentially decrease in time, starting from a finite value. There could instead be a relatively short ($< 1$ Gyr) transient period corresponding to the build-up phase with increasing SFR (D12), which is not captured by our simple assumptions. Thus a more complex version of the model (or numerical simulations) should be used if one is interested in the FMR at $z \simgt 6$. Secondly, to find an analytic solution to Eqs. 2-4, we have used an infall rate that is proportional to the SFR, although we do not make any assumption on its galaxy mass and redshift dependence. Thirdly, we have assumed that 30\% of the total gas mass in any galaxy is in the form of \HI, to be able to compare our results to the gas fraction observations of \citet{hughes2012}. In the future, we aim at exploring models wherein such assumptions are relaxed.

\section*{Acknowledgments} 
The authors thank G. Cresci, C. Evoli, F. Mannucci, R. Salvaterra and R. Schneider for useful discussions. JSD and PD acknowledge the support of the European Research Council via the award of an Advanced Grant, and JSD also acknowledges the support of the Royal Society via a Wolfson Research Merit award. PD thanks the SNS Pisa for their generous hospitality. Finally, the authors thank the anonymous referee for his/her insightful comments, which have added tremendously to the content of the paper.

 
\bibliographystyle{mn2e}
\bibliography{fr}

\newpage 
\label{lastpage} 
\end{document}